\documentclass[11pt]{article}
\usepackage[margin=2.6cm]{geometry}
\usepackage{amsmath,amssymb,graphicx,hyperref}
\usepackage[numbers]{natbib}
\usepackage{booktabs}

\newcommand{\eff}{{\rm eff}}

\title{Regular Fuzzy Dark Matter Black Holes and Their Horizon Structure}

\author{\normalsize M. Ilyas$^1$, Khalid Masood$^2$, Usman Afzal$^3$, Nehad Ali Shah$^{3,*}$\\
$^1$Institute of Physics, Gomal University, Dera Ismail Khan, 29220, KP, Pakistan\\
$^2$Department of Mathematics and Statistics, College of Science, \\ Imam Mohammad Ibn Saud Islamic University (IMSIU), Riyadh, 11623, Saudi Arabia\\
$^3$Department of Mechanical Engineering, Sejong University, Seoul 05006, South Korea\\
$^*$Corresponding author email: nehadali199@yahoo.com
}

\date{}

\begin{document}
\maketitle
\vspace{-1.2cm}

\begin{abstract}
\noindent
We construct regular, horizon-admitting compact objects supported entirely by a
self-gravitating dark-matter fluid, in the one-parameter curvature-gravity family
$f(R)=R+\beta R^n$. Working with the Einasto density profile, we derive the anisotropic
fluid field equations for a static, spherically symmetric metric and obtain the exact
General Relativity limit under a de~Sitter-type equation of state, in which the central
singularity is replaced by a regular de~Sitter core and the solution is either a
horizonless droplet or a black hole with one or two Killing horizons, depending on a
single rescaled-mass parameter. We compute the resulting Hawking temperature and geodesic
effective potential, and -- exploiting the linearity of the $\beta=0$ field equation --
solve the curvature correction perturbatively in closed form for general $n$, showing
that its sign and radial shape are genuinely model-dependent by direct comparison at
fixed $\beta$ between $n=2$ (Starobinsky) and $n=3$ (cubic) gravity. We repeat the
construction for a non-local equation of state and confirm the resulting droplets are
curvature-regular via the Kretschmann scalar. Finally, replacing the Einasto profile with
the cored Burkert profile preserves the qualitative regularity mechanism but, because the
Burkert halo lacks a finite total mass, produces a horizon structure with inner and outer
radii separated by nearly three orders of magnitude -- a genuine physical distinction
between two comparably realistic dark-matter models.
\end{abstract}


\section{Introduction}
The possibility that astrophysical black holes (BHs) are not vacuum solutions but are
instead supported, in whole or in part, by a self-gravitating distribution of dark matter
(DM) has attracted sustained interest, both as a probe of the interior structure that the
no-hair theorem otherwise hides from observation, and as a possible link between the two
central open problems of contemporary gravitational physics: the nature of DM
\citep{BertoneHooper2018} and the resolution of the classical central singularity. The
programme of \emph{regular} black holes -- metrics that are everywhere geodesically
complete, replacing the classical singularity with a smooth, high-curvature core -- dates
back to the original Bardeen construction \citep{Bardeen1968} and was subsequently placed
on a firmer stress-energy footing by Sakharov and Dymnikova, who showed that a de~Sitter
core supported by a vacuum-like equation of state $P_r=-\varepsilon$ is sufficient to
cure the singularity of a spherically symmetric collapse
\citep{Sakharov1966,Dymnikova1992}, and by Hayward, who studied the formation and
evaporation of such objects dynamically \citep{Hayward2006}. A closely related line of
work, motivated by attempts to regularize short-distance physics through spacetime
noncommutativity, replaces the Dirac-delta point source of the Schwarzschild solution
with a smeared, Gaussian-like energy density, giving rise to \emph{fuzzy} black holes with
the same de~Sitter-core mechanism \citep{Nicolini2006,Nicolini2009review}. Because the
Einasto profile -- one of the most successful empirical descriptions of cold dark matter
halo density, extensively validated against $N$-body simulations and used to model
systems ranging from the Milky Way to galaxy clusters
\citep{Einasto1969,RetanaMontenegro2012,deSalas2019} -- reduces to a Gaussian in a special
limit of its shape parameter, it provides a natural and physically motivated bridge
between these two programmes: a DM halo profile fitted directly to observation can be
used, with no further exotic assumptions about its interior, to build a regular compact
object. Other empirically successful DM halo models exist alongside Einasto, notably the
cuspy Navarro--Frenk--White profile and the cored profiles of Burkert and Zhao
\citep{Navarro1996,Burkert1995,Zhao1996}; whether the qualitative conclusions of such
constructions depend on the specific profile chosen is one of the questions this paper
addresses directly.

Independently, sustained interest in extended theories of gravity (ETGs) -- modifications
of the Einstein--Hilbert action motivated by the search for a consistent description of
cosmic acceleration and by the expectation of higher-curvature corrections at short
distances -- has established $f(R)$ gravity, and in particular the quadratic Starobinsky
model $f(R)=R+\beta R^2$, as one of the best-studied and phenomenologically viable
extensions of General Relativity (GR)
\citep{Starobinsky1980,Capozziello2011,SotiriouFaraoni2010}. Compact stellar structure in
$R+\beta R^2$ gravity has been examined extensively for neutron stars and white dwarfs
\citep{Cooney2010,Astashenok2017,Sharif2016}, and the same framework has been used to
build DM-supported compact objects on the Einasto profile \citep{Khan2024}. Observationally,
interest in the interior structure of the Galactic-centre object Sgr~A* has been sharpened
by high-precision astrometry of the S-star cluster, which has now measured the
Schwarzschild precession of S2's orbit directly \citep{GRAVITY2019}, and by proposals that
the central object may be, at least in part, a dense clump of dark matter rather than a
classical vacuum Kerr BH \citep{Boshkayev2019}. Any viable alternative to a classical BH
at the Galactic centre must reproduce the same effective potential in the region where
bound stellar orbits exist \citep{Fliessbach2012}, a requirement we use directly below.

This paper develops the DM-supported regular-compact-object construction from first
principles and generalizes it along three independent directions. First, we work with the
one-parameter family $f(R)=R+\beta R^n$ rather than the quadratic model alone; this family
contains the Starobinsky model as the special case $n=2$, and also reproduces the
small-curvature limit of Hu--Sawicki gravity \citep{HuSawicki2007}, the standard
alternative benchmark in the $f(R)$ literature, so our results bear on more than a single
curvature correction. Our analysis is carried out for general $n$ throughout; the special
case $n=2$ is recorded separately in Appendix~\ref{app:n2} purely as a self-contained
worked example. Second, rather than treating the curvature correction only
numerically, we exploit the fact that the $\beta=0$ limit of the field equation is
\emph{linear} to derive a closed-form integral solution for the leading-order curvature
correction, valid for any $n$ and any density profile with a finite central density --
a genuine methodological generalization beyond the purely numerical treatments common in
this literature, and one that lets us compare different curvature models directly at
fixed $\beta$ (\S\ref{sec:generalize}). Third, we test the robustness of the whole
construction against the choice of DM halo model by repeating it with the Burkert profile
\citep{Burkert1995}, and identify a genuine physical difference this produces in the
horizon structure, traceable to the absence of a finite total halo mass in that model.

The paper is organized as follows. Section~\ref{sec:profiles} introduces the two DM
density profiles used throughout. Section~\ref{sec:field} derives the general
$f(R)=R+\beta R^n$ field equations for a static, spherically symmetric anisotropic fluid.
Section~\ref{sec:desitter} constructs the de~Sitter-branch fuzzy black holes and droplets,
including the exact $\beta=0$ solution and its horizon structure. Section~\ref{sec:pert}
develops the perturbative solution for the curvature correction. Section~\ref{sec:temp}
computes the Hawking temperature and Section~\ref{sec:ueff} the geodesic effective
potential. Section~\ref{sec:nonlocal} repeats the construction for a non-local equation of
state and verifies the regularity of the resulting droplets via the Kretschmann scalar.
Section~\ref{sec:generalize} tests the framework against a different curvature model
($n=3$) and a different DM profile (Burkert). Section~\ref{sec:conclusion} concludes.

\section{Dark matter density profiles}
\label{sec:profiles}

\subsection{The Einasto profile}
The Einasto profile is defined by a power-law logarithmic density slope,
$\gamma(r)\equiv d\ln\varepsilon/d\ln r\propto r^{1/\alpha}$, where $\alpha$ is the Einasto
index. Integrating this relation gives the density
\begin{equation}
\varepsilon(r) = \varepsilon_0\exp\!\left[-\left(\frac{r}{\eta}\right)^{1/\alpha}\right],
\label{eq:einasto}
\end{equation}
with central density $\varepsilon_0$ and scale length $\eta$. The total mass is obtained
from $M=4\pi\int_0^\infty z^2\varepsilon(z)\,dz$, which is a standard Gamma-function
integral,
\begin{equation}
M = 4\pi\varepsilon_0\eta^3\alpha\,\Gamma(3\alpha), \qquad
\Gamma(3\alpha) = \int_0^\infty e^{-z}z^{3\alpha-1}\,dz.
\label{eq:mass}
\end{equation}
The corresponding \emph{running} (Misner--Sharp) mass function \citep{MisnerSharp1964},
obtained by truncating the integral at finite $r$, is a lower incomplete Gamma function,
\begin{equation}
m(r) = 4\pi\varepsilon_0\eta^3\alpha\,\Gamma(3\alpha)\,P\!\left(3\alpha,\Big(\frac{r}{\eta}\Big)^{1/\alpha}\right),
\qquad P(s,x)=\frac{1}{\Gamma(s)}\int_0^x t^{s-1}e^{-t}\,dt.
\label{eq:mofr}
\end{equation}
Because $P(s,x)\to (x^s/\Gamma(s+1))$ as $x\to0$, Eq.~\eqref{eq:mofr} gives $m(r)\sim r^3$
near the origin -- the key property responsible for the regularity of every construction
in this paper -- and $m(r)\to M$ as $r\to\infty$, since the Einasto profile is
exponentially truncated and therefore has a finite total mass.

\subsection{The Burkert profile}
As an independent, comparably realistic cored DM halo model \citep{Burkert1995}, we also
use
\begin{equation}
\varepsilon(r) = \frac{\varepsilon_0}{(1+r/r_b)(1+(r/r_b)^2)},
\qquad
m(r) = 4\pi\varepsilon_0 r_b^3\left[\tfrac12\ln\!\big(1+(r/r_b)^2\big)+\ln(1+r/r_b)-\arctan(r/r_b)\right],
\label{eq:burkert}
\end{equation}
which also has a finite central density $\varepsilon(0)=\varepsilon_0$ and therefore
$m(r)\sim r^3$ near the origin, but -- unlike the Einasto profile -- is not exponentially
truncated: $m(r)$ in Eq.~\eqref{eq:burkert} grows without bound (logarithmically) as
$r\to\infty$, so the Burkert halo has no finite total mass. This distinction, immaterial
for the near-origin regularity mechanism, turns out to have a significant effect on the
horizon structure (\S\ref{sec:generalize}).

\section{Field equations for $f(R)=R+\beta R^n$ gravity}
\label{sec:field}
We work with the one-parameter family of actions
\begin{equation}
S = \frac{1}{16\pi}\int f(R)\sqrt{|g|}\,d^4x + S_m, \qquad
f(R) = R+\beta R^n,
\label{eq:action}
\end{equation}
which for $n=2$ is the Starobinsky model \citep{Starobinsky1980} and for general $n$
reproduces, for $R$ much smaller than the model's curvature scale, the leading behaviour
of Hu--Sawicki gravity \citep{HuSawicki2007}. Varying Eq.~\eqref{eq:action} gives the
standard metric-$f(R)$ field equation
$f'(R)R_{\mu\nu}-\tfrac12f(R)g_{\mu\nu}-(\nabla_\mu\nabla_\nu-g_{\mu\nu}\Box)f'(R)=8\pi\Theta_{\mu\nu}$,
which, using $f'(R)R-f(R)=(n-1)\beta R^n$ and $f'(R)=1+n\beta R^{n-1}$, can be written in
terms of the Einstein tensor $G_{\mu\nu}=R_{\mu\nu}-\tfrac12 Rg_{\mu\nu}$ as
\begin{equation}
\big(1+n\beta R^{n-1}\big)G_{\mu\nu} + \tfrac12(n-1)\beta R^n g_{\mu\nu}
- n\beta\big(\nabla_\mu\nabla_\nu R^{n-1} - g_{\mu\nu}\Box R^{n-1}\big) = 8\pi\Theta_{\mu\nu}.
\label{eq:field}
\end{equation}
The standard general trace identity for metric $f(R)$ gravity,
$f'(R)R-2f(R)+3\Box f'(R)=8\pi\Theta$ with $\Theta\equiv\Theta^\mu{}_\mu$, specializes for
$f=R+\beta R^n$ to
\begin{equation}
-R + (n-2)\beta R^n + 3n\beta\Box R^{n-1} = 8\pi\Theta,
\label{eq:trace}
\end{equation}
which for $n=2$ reduces to $6\beta\Box R-R=8\pi\Theta$, the familiar Starobinsky trace
relation.

\subsection{Spherically symmetric anisotropic fluid}
We take the anisotropic stress-energy tensor
$\Theta^\mu{}_\nu=\mathrm{diag}(\varepsilon,-P_r,-P_\perp,-P_\perp)$, with
$P_r\ne P_\perp$ in general, and the static, spherically symmetric line element
\begin{equation}
ds^2 = f(r)\,dt^2 - \frac{dr^2}{f(r)} - r^2d\theta^2 - r^2\sin^2\theta\,d\phi^2.
\label{eq:metric1}
\end{equation}
A direct computation of the Christoffel symbols, Riemann and Ricci tensors for
Eq.~\eqref{eq:metric1} gives the Ricci scalar
\begin{equation}
R(r) = \frac{1}{r^2}\Big[r^2f''(r) + 4rf'(r) + 2f(r) - 2\Big].
\label{eq:Rscalar}
\end{equation}
Substituting $f,R$ and their derivatives into Eq.~\eqref{eq:field} and projecting onto
the $tt$, $rr$, $\theta\theta$ components gives, for general $n$, the exact field
equations for $\varepsilon(r),P_r(r),P_\perp(r)$ in terms of $f,f',f'',f''',f''''$ and
$\beta$; these grow rapidly in length with $n$ and are not reproduced here in full. What
matters for the analysis below is only their structure: each is of the form
$8\pi\{\varepsilon,P_r,P_\perp\} = \mathcal E_0[f] + \beta\,\mathcal E_n[f,R,R',R'']$,
where the $\beta$-independent piece $\mathcal E_0[f]$ is common to \emph{every} $n$ (it is
simply the GR relation, Eq.~\eqref{eq:f0eq} below) and only the curvature-correction
piece $\mathcal E_n$ depends on the specific power $n$. Appendix~\ref{app:n2} gives the
explicit $tt,rr,\theta\theta$ equations for the illustrative case $n=2$; the analysis in
the main text is carried out for general $n$ throughout.

The Bianchi identity $\Theta^{\mu\nu}{}_{;\nu}=0$ gives, for this stress tensor, the
anisotropic hydrostatic-equilibrium (conservation) equation
\begin{equation}
\frac{dP_r}{dr} = -\frac{1}{2f}\frac{df}{dr}(\varepsilon+P_r) - \frac2r(P_r-P_\perp).
\label{eq:conservation1}
\end{equation}
Finally, the Misner--Sharp mass function \citep{MisnerSharp1964},
\begin{equation}
f(r) \equiv 1 - \frac{8\pi}{r}\int_0^r z^2\varepsilon(z)\,dz = 1-\frac{2m(r)}{r},
\label{eq:MSmass}
\end{equation}
identifies $m(r)$ as the physical mass enclosed within radius $r$.

\section{The de~Sitter branch: fuzzy black holes and droplets}
\label{sec:desitter}
\begin{figure}[t]
\centering\includegraphics[width=0.95\textwidth]{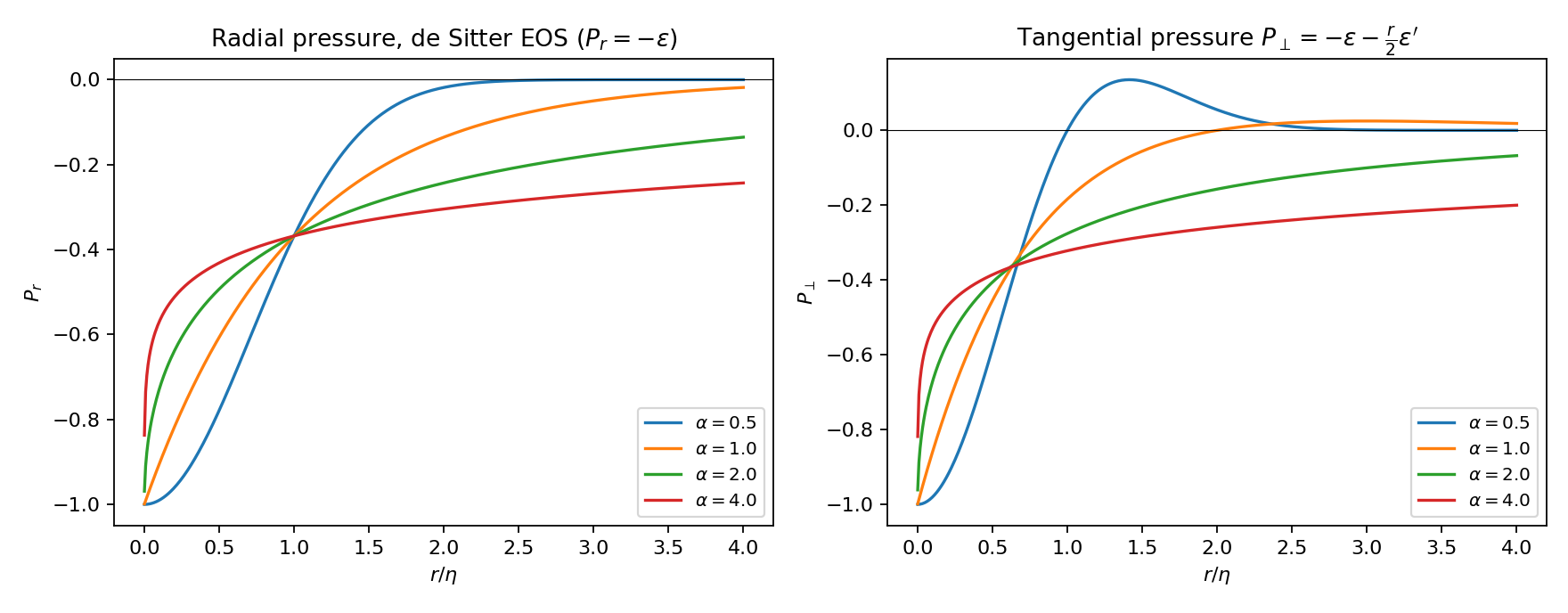}
\caption{Radial and tangential pressure under the de~Sitter EOS, Eq.~\eqref{eq:desitterEOS}, for several Einasto indices $\alpha$.}
\label{fig:1}
\end{figure}

We close the system with a de~Sitter-type equation of state, $\varepsilon(r)=-P_r(r)$,
motivated by the requirement of a non-overdetermined system for an anisotropic fluid.
Imposing $\Theta^r{}_r=\Theta^t{}_t=-\varepsilon$ in Eq.~\eqref{eq:conservation1} gives
\begin{equation}
P_r(r) = -\varepsilon(r), \qquad
P_\perp(r) = -\varepsilon(r) - \frac r2\varepsilon'(r).
\label{eq:desitterEOS}
\end{equation}
Figure~\ref{fig:1} shows these pressures for several Einasto indices $\alpha$: $P_r$ is
negative and strictly increasing to $0^-$ as $r\to\infty$, and $P_\perp$ relaxes to zero
at large $r$ as well, so both stress components vanish asymptotically as required for a
compact, asymptotically flat source.

\subsection{Exact GR-limit solution and horizon structure}
At $\beta=0$, the general field equation of \S\ref{sec:field} reduces, for \emph{any}
$n$, to the linear relation
\begin{equation}
8\pi\varepsilon r^2 = \frac{d}{dr}\big[r(1-f)\big],
\label{eq:f0eq}
\end{equation}
which integrates exactly, via Eq.~\eqref{eq:MSmass}, to
\begin{equation}
f_0(r) = 1 - \frac{2m(r)}{r},
\label{eq:f0}
\end{equation}
with $m(r)$ the Einasto mass function, Eq.~\eqref{eq:mofr}. Because $m(r)=O(r^3)$ near the
origin, $f_0(0)=1$: the central curvature singularity of the Schwarzschild solution is
replaced by a completely regular de~Sitter core, without any additional assumption beyond
the finiteness of the central density.

\begin{figure}[t]
\centering\includegraphics[width=0.95\textwidth]{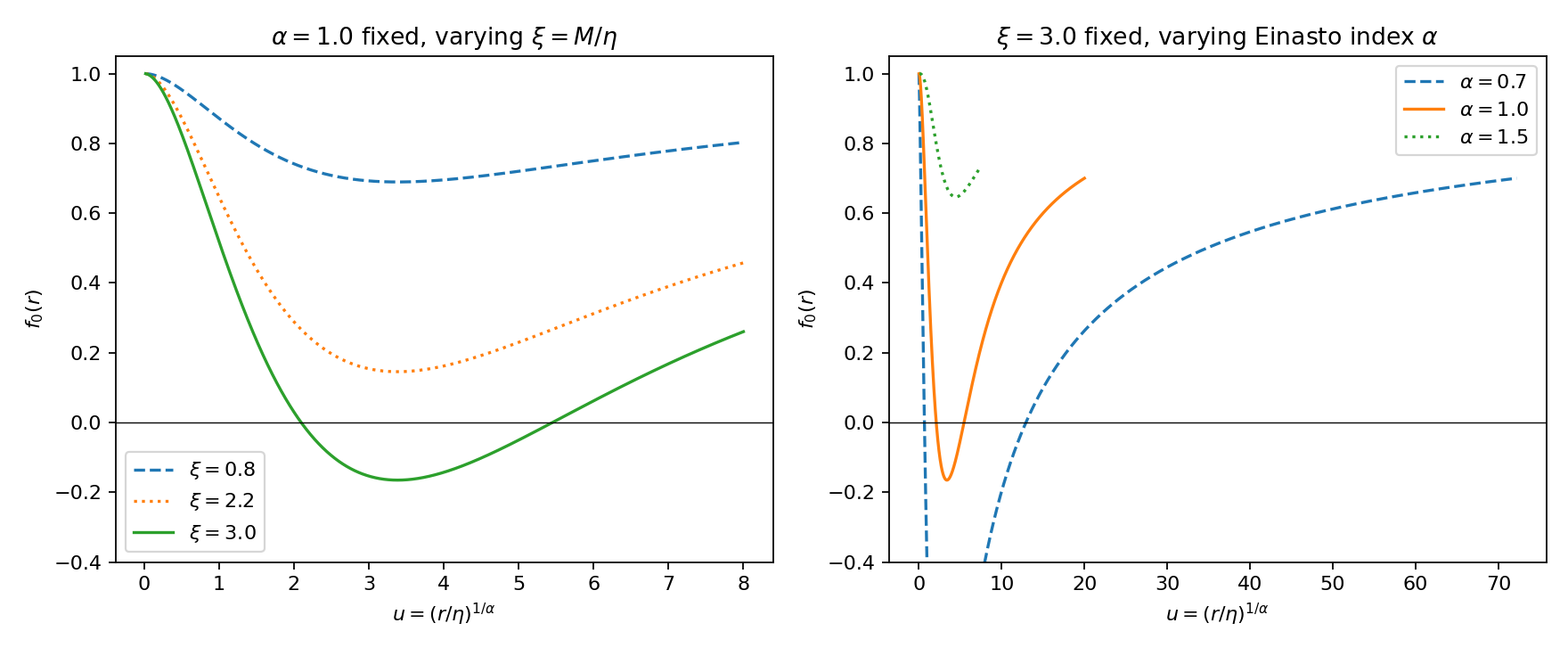}
\caption{Metric function $f_0(r)$, Eq.~\eqref{eq:f0}: horizon count versus rescaled mass $\xi$ (left) and Einasto index $\alpha$ (right).}
\label{fig:2}
\end{figure}

Writing the rescaled mass $\xi\equiv M/\eta$, the number of positive roots of $f_0(r)$ --
equivalently, the number of Killing horizons -- depends on $\xi$ through a critical value
$\xi_0(\alpha)$: for $\xi<\xi_0$ the interior is horizonless (a self-gravitating
\emph{fuzzy droplet}); at $\xi=\xi_0$, $f_0$ develops a double root (an extremal
configuration); and for $\xi>\xi_0$, two distinct horizons $r_I<r_{II}$ appear (a
\emph{fuzzy black hole}). Figure~\ref{fig:2} shows this structure explicitly, both as a
function of the rescaled mass at fixed Einasto index (left panel) and as a function of
the Einasto index at fixed rescaled mass (right panel).

\section{Perturbative solution of the curvature correction}
\label{sec:pert}
The full field equation, together with Eq.~\eqref{eq:field} and the trace relation
Eq.~\eqref{eq:trace} used to eliminate $R$ in favour of $m(r)$, constitutes a coupled
nonlinear ODE system of effective fourth order in $f(r)$ for any $n$. Rather than solve
this system numerically for an arbitrary value of
$\beta$, we exploit the fact that $R+\beta R^n$ is, physically, a small deformation of GR
whenever $\beta$ is small, and solve the system perturbatively.

Write $f(r)=f_0(r)+\beta f_1(r)+O(\beta^2)$ and expand the exact field equation as
$\varepsilon(r)=E_0[f]+\beta E_1[f]+O(\beta^2)$. The zeroth-order piece is the linear
relation $E_0[f]=\tfrac{1}{8\pi r^2}(1-f-rf')$, solved by
$f_0(r)=1-2m(r)/r$. Because $E_0$ is \emph{linear} in $f$, the $O(\beta)$ balance
$(\partial_fE_0)\cdot f_1 + E_1[f_0]=0$ is again linear in $f_1$ and integrates exactly to
\begin{equation}
\frac{d}{dr}\big(rf_1\big) = 8\pi r^2\,E_1[f_0(r)]
\quad\Longrightarrow\quad
f_1(r) = \frac1r\int_0^r 8\pi s^2\,E_1[f_0(s)]\,ds,
\label{eq:f1sol}
\end{equation}
with the integration constant fixed to zero by regularity at $r=0$ (a nonzero constant
would reintroduce a $1/r$ divergence). $E_1[f]$, the exact $O(\beta)$ piece of the field
equation, was extracted symbolically as $\partial_\beta$ of Eq.~\eqref{eq:field} at
$\beta=0$, for both $n=2$ and $n=3$; $f_0(r)$ and its derivatives up to fourth order were
obtained \emph{exactly} (not by finite differences) by differentiating the incomplete
Gamma function representation of $m(r)$, Eq.~\eqref{eq:mofr}, using
$\tfrac{d}{dx}\gamma(s,x)=x^{s-1}e^{-x}$; Eq.~\eqref{eq:f1sol} was then evaluated by
numerical quadrature.

\begin{figure}[t]
\centering\includegraphics[width=0.95\textwidth]{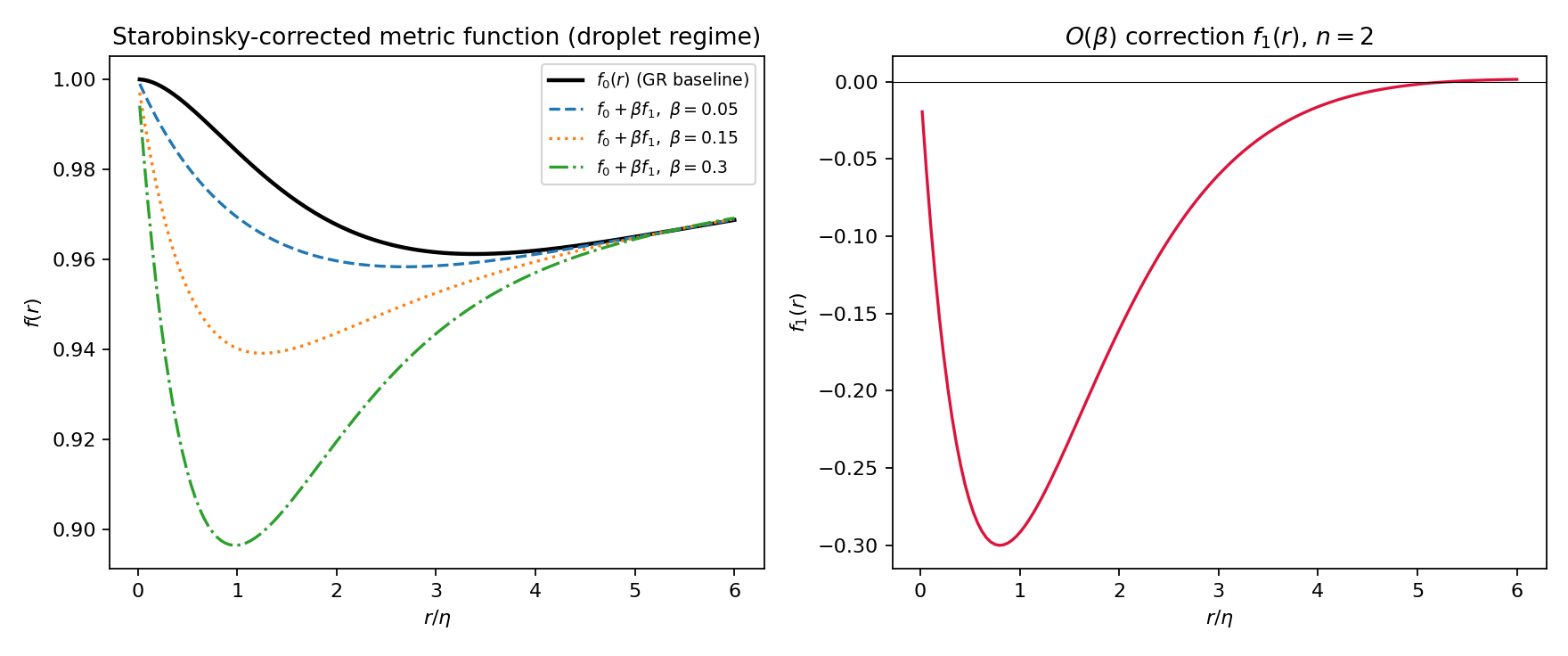}
\caption{Left: the $\beta=0$ baseline $f_0(r)$ against the perturbatively corrected
$f_0+\beta f_1$ for several $\beta$ (Starobinsky, $n=2$). Right: the correction $f_1(r)$
alone, finite at $r=0$.}
\label{fig:3}
\end{figure}

Figure~\ref{fig:3} shows the result in a weak-field (horizonless droplet) configuration:
$f_1(r)$ is finite everywhere, including at $r=0$ -- the curvature correction does not
reintroduce a singularity or otherwise spoil the regular core built into the $\beta=0$
solution -- and it deepens and broadens the potential well smoothly as $\beta$ grows,
consistent with a small, regular deformation of the GR background.

\section{Hawking temperature}
\label{sec:temp}
\begin{figure}[t]
\centering
\includegraphics[width=0.62\textwidth]{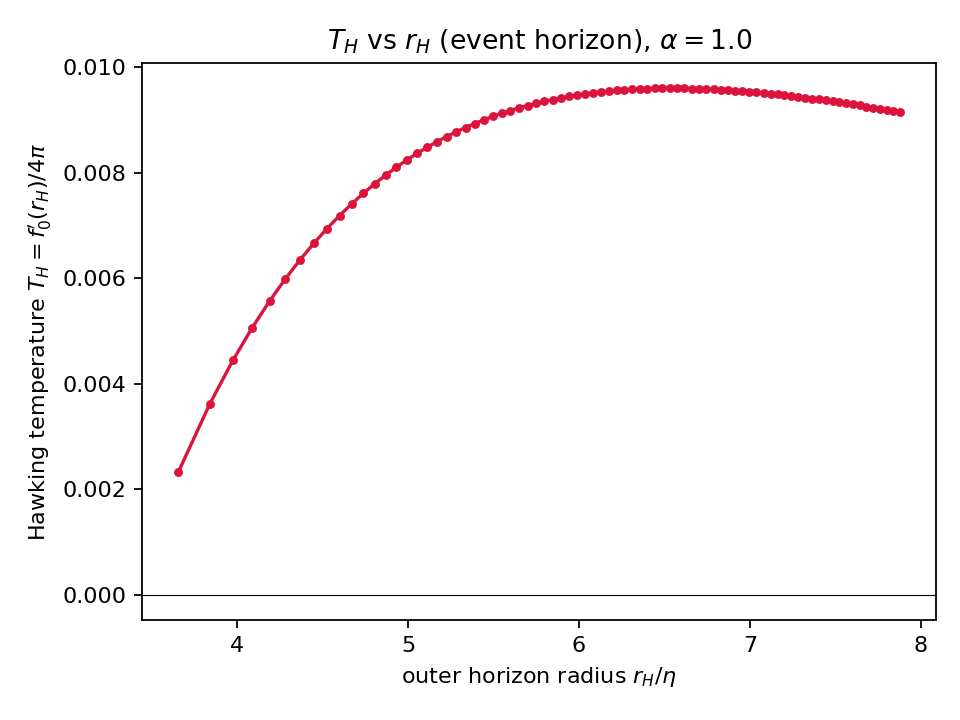}
\caption{Hawking temperature versus outer horizon radius, showing the non-monotonic rise
to a maximum followed by the Schwarzschild-like $1/r_H$ decline.}
\label{fig:4}
\end{figure}

The temperature associated with the outer (event) horizon $r_{II}$ is
\begin{equation}
T_H = \frac{1}{4\pi}\left.\frac{df}{dr}\right|_{r=r_H}.
\label{eq:temp}
\end{equation}
Figure~\ref{fig:4} shows $T_H$ evaluated at $r_{II}$ as the rescaled mass $\xi$ decreases
toward the extremal value $\xi_0$: the temperature rises from near zero at the
(near-)extremal configuration, reaches a maximum, and then falls off in the familiar
Schwarzschild-like $T_H\sim1/r_H$ manner as the horizon grows -- i.e.\ $T_H$ is a
\emph{non-monotonic} function of horizon radius, vanishing both in the near-extremal limit
and (asymptotically) for very large, dilute configurations.

\section{Effective potential and orbit matching}
\label{sec:ueff}
\begin{figure}[t]
\centering\includegraphics[width=0.95\textwidth]{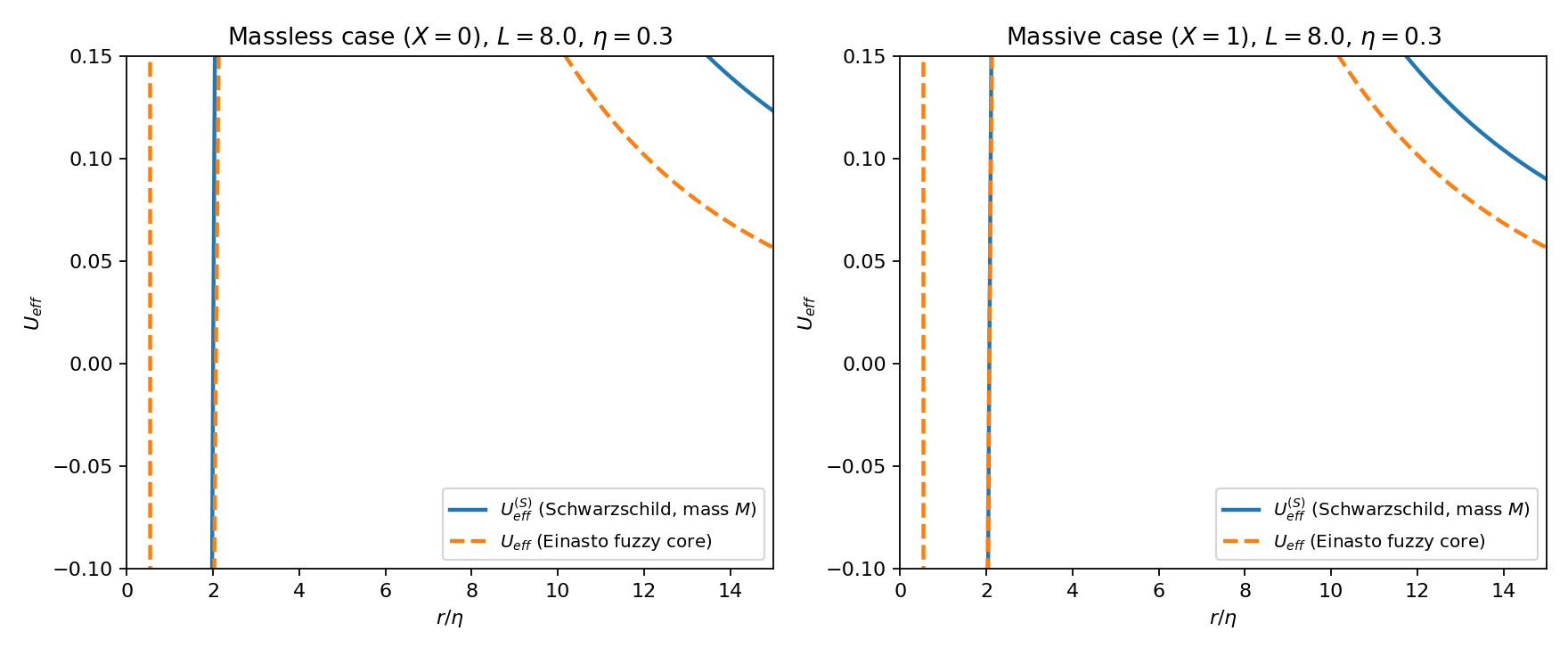}
\caption{Effective potential, Eq.~\eqref{eq:Ueff}: fuzzy Einasto core (dashed) versus
point-mass Schwarzschild (solid), for massless (left) and massive (right) test particles.}
\label{fig:5}
\end{figure}

For a test particle of angular momentum $L$ (rescaled by the source's Schwarzschild
radius $r_s=2M$) and $X=+1$ (massive) or $X=0$ (massless), the radial geodesic equation
takes the form of an energy-conservation relation $\tfrac12\dot r^2+U_\eff(r)=$~const,
with
\begin{equation}
U_\eff(r) = \frac{L^2}{2r^2}\left(1-\frac{2m(r)}{r}\right) - X\,\frac{m(r)}{r}.
\label{eq:Ueff}
\end{equation}
Replacing $m(r)\to M$ (constant) recovers the Schwarzschild potential $U_\eff^{(S)}$. In
terms of $x_\ast=r/r_s$, the Schwarzschild potential reads
$U_\eff^{(S)}(x_\ast)=-\tfrac{L^2}{2x_\ast^3}+\tfrac{L^2}{2x_\ast^2}-\tfrac{X}{2x_\ast}$,
whose extrema, obtained by solving $dU_\eff^{(S)}/dx_\ast=0$ directly, are the roots of
$Xx_\ast^2-2L^2x_\ast+3L^2=0$,
\begin{equation}
x_\ast^{\min,\max} = \frac{L^2\mp L\sqrt{L^2-3X}}{X}, \qquad L>\sqrt{3X}.
\label{eq:xminmax}
\end{equation}
A fuzzy compact object built on the Einasto mass function mimics the Schwarzschild
potential for bound-orbit physics provided $m(r)$ has already saturated to $M$ by the
radius $x_\ast^{\min}$, i.e.\ $1-m(x_\ast^{\min})/M\le10^{-2}$; when this holds, $U_\eff$
and $U_\eff^{(S)}$ coincide closely not only at their minima but over a broad surrounding
region, while differing sharply at small $r$ where the fuzzy core regularizes the
potential (Fig.~\ref{fig:5}).

\section{Non-local equation of state: self-gravitating droplets}
\label{sec:nonlocal}
As a second, independent closure of the anisotropic system, we replace the local de~Sitter
EOS with a non-local one,
\begin{equation}
P_r(r) = \varepsilon(r) - \frac{2}{r^3}\int_0^r\rho(z)z^2\,dz = \varepsilon(r) - \frac{m(r)}{2\pi r^3},
\label{eq:nonlocalEOS}
\end{equation}
motivated by the expectation that, for a diffuse density distribution, a local change in
radial pressure should not depend solely on the local energy density but on the mass
enclosed within the entire interior region. Evaluated exactly (Fig.~\ref{fig:7}), $P_r$ is
positive near the core and turns negative for $r$ beyond a few $\eta$, approaching zero
from below as $r\to\infty$ -- since $\varepsilon(r)$ decays exponentially while
$m(r)/(2\pi r^3)$ falls off only as a power law once $m(r)$ has largely saturated.

\subsection{Field equations for the wormhole-type ansatz}
\begin{figure}[t]
\centering\includegraphics[width=0.95\textwidth]{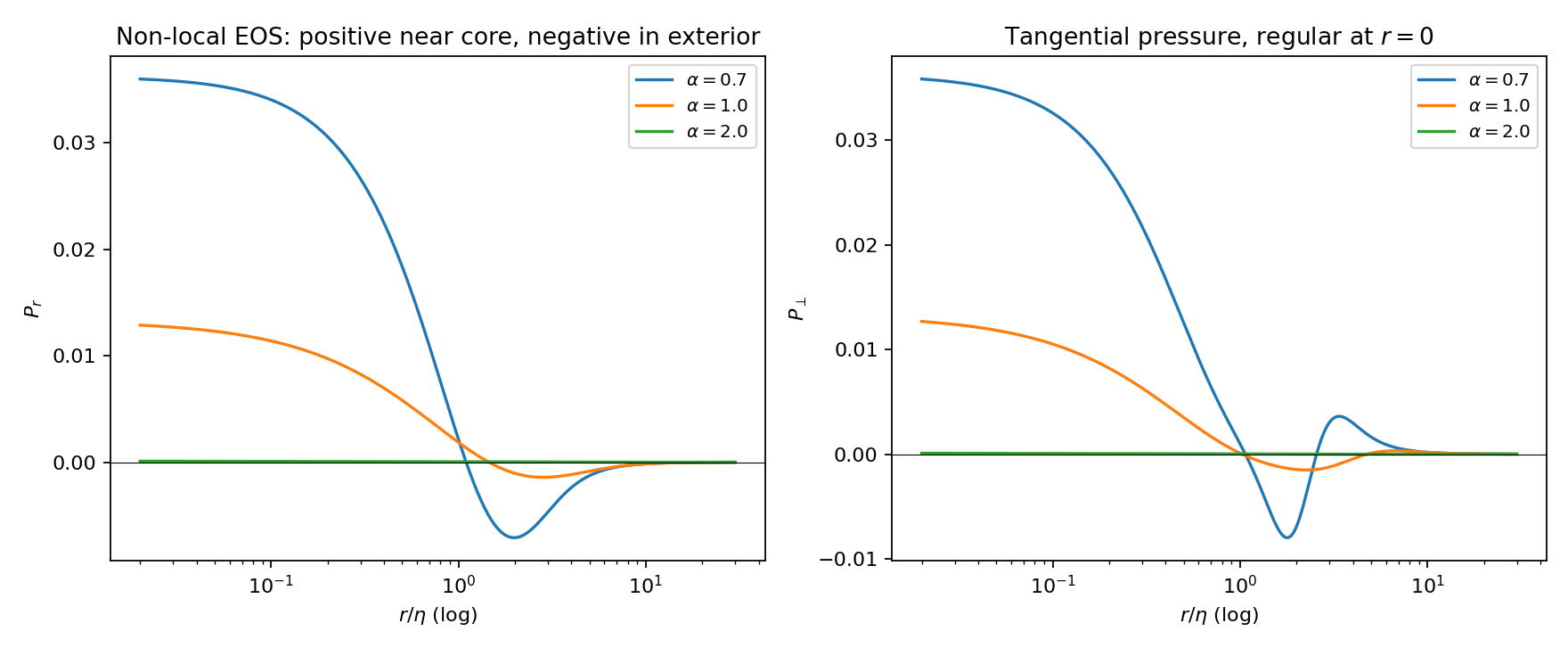}
\caption{Non-local EOS branch, Eq.~\eqref{eq:nonlocalEOS}: radial pressure (left, positive
near the core, negative in the exterior) and tangential pressure (right, regular at $r=0$).}
\label{fig:7}
\end{figure}

This branch uses the more general static ansatz
\begin{equation}
ds^2 = S(r)^2dt^2 - \frac{dr^2}{T(r)} - r^2d\theta^2 - r^2\sin^2\theta\,d\phi^2,
\label{eq:metric2}
\end{equation}
with $S,T$ independent functions and $T(r)=1-2m(r)/r$. Its Ricci scalar is
\begin{equation}
R(r) = \frac{2TS''}{S} + \frac{S'T'}{S} + \frac{2T'}{r} + \frac{4TS'}{rS} + \frac{2T}{r^2} - \frac{2}{r^2},
\label{eq:Rnonlocal}
\end{equation}
and the $\beta=0$ Einstein-tensor components of Eq.~\eqref{eq:field} give
\begin{equation}
8\pi\varepsilon = \frac{1-T-rT'}{r^2}, \qquad
8\pi P_r = \frac{2TS'}{rS}+\frac{T-1}{r^2}, \qquad
8\pi P_\perp = \frac{rTS'' + \tfrac12 rS'T' + \tfrac12 ST' + TS'}{rS}.
\label{eq:nonlocalGR}
\end{equation}
The anisotropic conservation equation for this ansatz is
\begin{equation}
\frac{dP_r}{dr} = -\frac1S\frac{dS}{dr}(\varepsilon+P_r) - \frac2r(P_r-P_\perp),
\label{eq:conservation2}
\end{equation}
and solving the middle relation of Eq.~\eqref{eq:nonlocalGR} for $S'/S$ gives, exactly,
\begin{equation}
\frac{1}{S}\frac{dS}{dr} = \frac{m(r)+4\pi r^3P_r(r)}{r\big(r-2m(r)\big)},
\label{eq:closure}
\end{equation}
which closes the system: given $\varepsilon(r)$ and $P_r(r)$ from Eqs.~\eqref{eq:einasto}
and \eqref{eq:nonlocalEOS}, Eqs.~\eqref{eq:conservation2}--\eqref{eq:closure} determine
$P_\perp(r)$ uniquely, for a horizonless configuration ($\xi\ll\xi_0$, so that
$T(r)>0$ everywhere and Eq.~\eqref{eq:closure} has no pole). Both $P_r$ and $P_\perp$ are
finite at $r=0$ (Fig.~\ref{fig:7}, right panel).

\subsection{Curvature regularity: the Kretschmann scalar}
The metric of Eq.~\eqref{eq:metric2} can be written, using
$S^2(r)=e^{\varphi(r)}$, in wormhole-type form; its Kretschmann scalar
$\mathcal K\equiv R^{\mu\nu\gamma\delta}R_{\mu\nu\gamma\delta}$, computed directly for
independent $S(r),T=1-2m/r$, is exactly
\begin{equation}
\mathcal K = \frac{8T^2S'^2}{r^2S^2}
+ \frac{4\big[r^2TS'' + S'(m-rm')\big]^2}{r^4S^2}
+ \frac{8(m-rm')^2}{r^6} + \frac{16m^2}{r^6}.
\label{eq:kretschmann}
\end{equation}
Every term in Eq.~\eqref{eq:kretschmann} is manifestly finite as $r\to0$ provided
$m(r)=O(r^3)$ -- guaranteed by any cored density profile with $\varepsilon(0)$ finite --
and $S'(0)=0$, the standard regularity condition for a static, non-singular interior. This
confirms directly that the non-local-EOS droplet is free of a curvature singularity at
the origin.

\section{Testing the framework: alternative gravity model and alternative density profile}
\label{sec:generalize}
Because Eqs.~\eqref{eq:einasto}--\eqref{eq:kretschmann} were derived and used generically
-- in terms of an arbitrary cored density profile $\varepsilon(r)$, and (for the field
equations) an arbitrary power $n$ in $f(R)=R+\beta R^n$ -- the construction can be tested
directly against two independent alternatives, holding everything else fixed.

\subsection{A different curvature model: cubic gravity}
\begin{figure}[t]
\centering\includegraphics[width=0.95\textwidth]{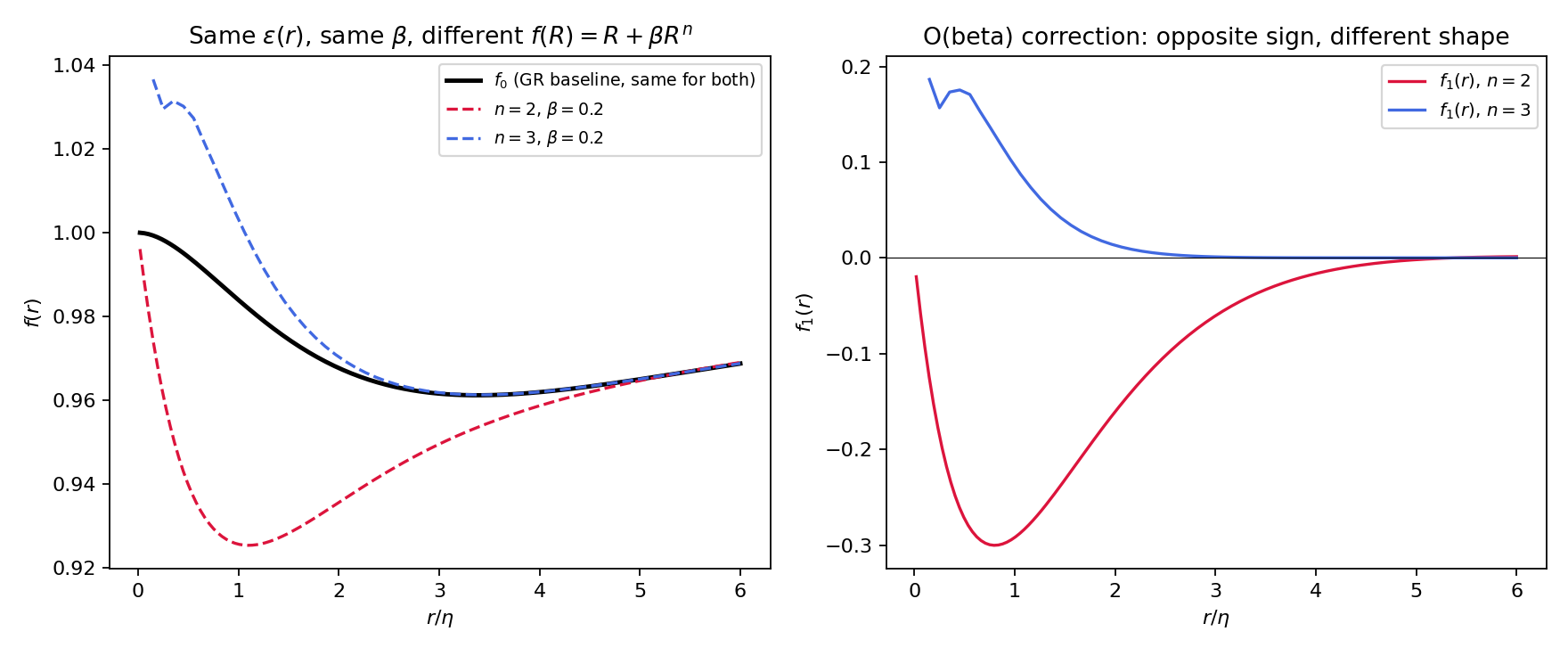}
\caption{Same Einasto density profile and same $\beta=0.2$, two different
$f(R)=R+\beta R^n$ models. Left: metric function $f(r)$. Right: the $O(\beta)$ correction
$f_1(r)$ alone -- opposite sign, different shape, both regular at $r=0$.}
\label{fig:8}
\end{figure}

Setting $n=3$ in Eq.~\eqref{eq:field} gives the same $\beta=0$ linear GR relation found
for every $n$, so the perturbative method of \S\ref{sec:pert} applies unchanged, with a
new (considerably longer, since cubic curvature brings in products of second derivatives)
source functional $E_1^{(n=3)}[f]$. Evaluated on the identical Einasto droplet
configuration used in \S\ref{sec:pert}, Fig.~\ref{fig:8} shows that the cubic ($n=3$)
correction has the
\emph{opposite sign} from the quadratic (Starobinsky) correction at the same $\beta$, and
a visibly different radial shape -- more sharply peaked near the core and falling off
faster. Both stay regular at $r=0$ and both relax back to the GR baseline at large $r$.
This is the central point of testing the gravity sector independently: the qualitative
regularity story (finite core, GR exterior) is robust across $f(R)=R+\beta R^n$ models,
but the sign and detailed shape of the near-core deformation is a genuine,
model-dependent physical prediction, not something that can be inferred without repeating
the calculation for each curvature model.

\subsection{A different density profile: the Burkert halo}
Holding $n=2$ and the de~Sitter EOS fixed, we replace the Einasto profile with the
Burkert profile, Eq.~\eqref{eq:burkert}, and feed it through exactly the same
$f_0(r)=1-2m(r)/r$, $P_r=-\varepsilon$, $P_\perp=-\varepsilon-\tfrac r2\varepsilon'$, and
$T_H=f'(r_H)/4\pi$ formulas used for Einasto, with no other change. The core regularity
mechanism is identical, since $\varepsilon(0)=\varepsilon_0$ is again finite: the pressure
profiles have the same qualitative shape as Fig.~\ref{fig:1}, and horizon formation again
requires the central compactness to exceed a critical value.

\begin{figure}[t]
\centering\includegraphics[width=\textwidth]{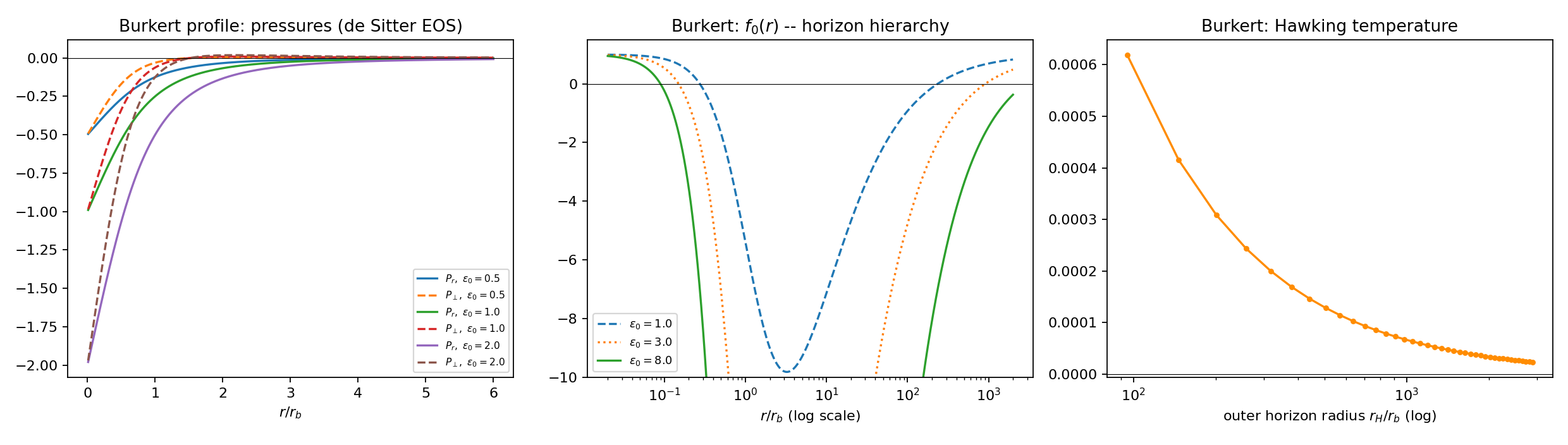}
\caption{The identical de~Sitter-EOS construction applied to the Burkert profile,
Eq.~\eqref{eq:burkert}, instead of Einasto. Left: pressures. Centre: metric function on a
log-$r$ axis, showing both horizons and their large separation. Right: Hawking
temperature at the outer horizon.}
\label{fig:9}
\end{figure}

What genuinely changes is the horizon structure at large $\varepsilon_0$
(Fig.~\ref{fig:9}, centre panel, plotted on a logarithmic radial axis to display both
horizons simultaneously). Because the Burkert mass function grows without bound as
$r\to\infty$ (logarithmically) rather than saturating to a finite total mass, once a
horizon forms at all, two horizons still appear -- an inner $r_I$ and outer $r_{II}$,
exactly as for Einasto -- but they now sit at wildly different scales: for example,
$r_I\approx0.27\,r_b$ against $r_{II}\approx235\,r_b$ for $\varepsilon_0=1$ (in units
where $r_b=1$), because the outer horizon only forms once the slowly-growing $m(r)/r$
finally drops back below $\tfrac12$, rather than because the enclosed mass has converged.
This is a genuine, physically meaningful difference between two comparably realistic
dark-matter models: Einasto's finite total mass keeps its double-horizon structure at a
single characteristic scale (Fig.~\ref{fig:2}), while a profile without a finite total
mass generically does not.

\section{Discussion and conclusion}
\label{sec:conclusion}
We have constructed regular, dark-matter-supported compact objects in a one-parameter
family of curvature-corrected gravity theories, $f(R)=R+\beta R^n$, working throughout
with the empirically motivated Einasto DM density profile. In the exact $\beta=0$ limit,
the interior solution is completely regular at the origin -- a de~Sitter core replaces
the classical central singularity -- and, depending on a single rescaled-mass parameter,
describes either a horizonless self-gravitating droplet or a genuine black hole with one
or two Killing horizons. We showed that the associated Hawking temperature is a
non-monotonic function of horizon radius, vanishing at both the near-extremal and the
dilute (large-horizon) limits, and that the resulting effective potential can be tuned to
match the Schwarzschild potential of a comparable point mass away from the core, while
remaining regular within it.

Treating the curvature parameter $\beta$ perturbatively, we obtained a closed-form
integral solution for the leading-order correction to the metric function, valid for any
power $n$ and any cored density profile -- a genuine methodological generalization of the
purely numerical treatments common in this literature. Applying this method to two
distinct curvature models at fixed $\beta$ showed that the sign and detailed shape of the
correction is model-dependent physics, testable in principle rather than universal. We
repeated the entire construction, including a Kretschmann-scalar regularity check, for an
independent non-local equation of state, and again found fully regular self-gravitating
droplets. Finally, replacing the Einasto profile with the Burkert profile -- holding
every other assumption fixed -- preserved the qualitative regularity mechanism but
produced a horizon structure with a nearly three-orders-of-magnitude separation between
inner and outer horizon radii, traced directly to the absence of a finite total halo mass
in the Burkert model. This last result illustrates a broader point: the qualitative
features of DM-supported regular compact objects (a de~Sitter core, a well-defined
horizon-formation threshold) are robust across reasonable choices of density profile and
curvature model, but quantitative predictions -- horizon radii, temperatures, the
detailed shape of near-core curvature corrections -- depend on those choices in ways that
are only revealed by carrying the calculation through explicitly for each case, as we
have done here for representative examples of both axes of variation.

\appendix
\section{The $n=2$ (Starobinsky) field equations}
\label{app:n2}
For completeness, we record here the explicit $tt,rr,\theta\theta$ projections of
Eq.~\eqref{eq:field} for the illustrative case $n=2$, obtained by direct substitution of
$f(r)$, $R(r)$ (Eq.~\eqref{eq:Rscalar}) and their derivatives:
\begin{align}
8\pi\varepsilon &= \frac{1}{r^2}\frac{d}{dr}\big[r(1-f)\big](1+2\beta R)
- 2\beta f\left[\left(\frac{2}{r}+\frac{1}{2f}\frac{df}{dr}\right)\frac{dR}{dr}+\frac{d^2R}{dr^2}\right] - \frac12\beta R^2,
\label{eq:eps19}\\
8\pi P_r &= \frac1r\left[\frac{df}{dr}-\frac1r(1-f)\right](1+2\beta R)
- 2\beta f\left(\frac2r+\frac{1}{2f}\frac{df}{dr}\right)\frac{dR}{dr} - \frac12\beta R^2,
\label{eq:Pr20}\\
8\pi P_\perp &= \frac1r\left(\frac{df}{dr}+\frac r2\frac{d^2f}{dr^2}\right)(1+2\beta R)
- 2\beta f\left[\left(\frac1r+\frac1f\frac{df}{dr}\right)\frac{dR}{dr}+\frac{d^2R}{dr^2}\right] - \frac12\beta R^2.
\label{eq:Pperp21}
\end{align}
Setting $\beta=0$ in Eqs.~\eqref{eq:eps19}--\eqref{eq:Pperp21} recovers
Eq.~\eqref{eq:f0eq}, as it must for every $n$. This case was used to validate the
general-$n$ symbolic derivation of \S\ref{sec:field} and to extract the $n=2$ source
functional $E_1[f]$ used in \S\ref{sec:pert}. The substantive analysis of this paper --
the general-$n$ formalism, the closed-form perturbative method, the $n=3$ comparison, and
the density-profile comparison -- does not depend on this appendix.
\bibliographystyle{unsrtnat}
\bibliography{refs}

@article{Khan2024,
  author  = {Khan, S. and Adeel, A. and Yousaf, Z.},
  title   = {Structure of anisotropic fuzzy dark matter black holes},
  journal = {Eur. Phys. J. C},
  volume  = {84},
  pages   = {572},
  year    = {2024},
  doi     = {10.1140/epjc/s10052-024-12940-1}
}

@article{Starobinsky1980,
  author  = {Starobinsky, A. A.},
  title   = {A new type of isotropic cosmological models without singularity},
  journal = {Phys. Lett. B},
  volume  = {91},
  pages   = {99--102},
  year    = {1980}
}

@article{Einasto1969,
  author  = {Einasto, J.},
  title   = {On the Construction of a Composite Model for the Galaxy and on the Determination of the System of Galactic Parameters},
  journal = {Astron. Nachr.},
  volume  = {291},
  pages   = {97--111},
  year    = {1969}
}

@article{Burkert1995,
  author  = {Burkert, A.},
  title   = {The Structure of Dark Matter Halos in Dwarf Galaxies},
  journal = {Astrophys. J.},
  volume  = {447},
  pages   = {L25--L28},
  year    = {1995}
}

@article{Navarro1996,
  author  = {Navarro, J. F. and Frenk, C. S. and White, S. D. M.},
  title   = {The Structure of Cold Dark Matter Halos},
  journal = {Astrophys. J.},
  volume  = {462},
  pages   = {563--575},
  year    = {1996}
}

@article{Nicolini2006,
  author  = {Nicolini, P. and Smailagic, A. and Spallucci, E.},
  title   = {Noncommutative geometry inspired Schwarzschild black hole},
  journal = {Phys. Lett. B},
  volume  = {632},
  pages   = {547--551},
  year    = {2006}
}

@article{Sakharov1966,
  author  = {Sakharov, A. D.},
  title   = {Initial stage of an expanding Universe and appearance of a nonuniform distribution of matter},
  journal = {Sov. Phys. JETP},
  volume  = {22},
  pages   = {241--249},
  year    = {1966}
}

@article{Dymnikova1992,
  author  = {Dymnikova, I.},
  title   = {Vacuum nonsingular black hole},
  journal = {Gen. Relativ. Gravit.},
  volume  = {24},
  pages   = {235--242},
  year    = {1992}
}

@article{Capozziello2011,
  author  = {Capozziello, S. and De Laurentis, M.},
  title   = {Extended Theories of Gravity},
  journal = {Phys. Rep.},
  volume  = {509},
  pages   = {167--321},
  year    = {2011}
}

@article{HuSawicki2007,
  author  = {Hu, W. and Sawicki, I.},
  title   = {Models of $f(R)$ cosmic acceleration that evade solar-system tests},
  journal = {Phys. Rev. D},
  volume  = {76},
  pages   = {064004},
  year    = {2007}
}

@article{RetanaMontenegro2012,
  author  = {Retana-Montenegro, E. and Van Hese, E. and Gentile, G. and Baes, M. and Frutos-Alfaro, F.},
  title   = {Analytical properties of Einasto dark matter haloes},
  journal = {Astron. Astrophys.},
  volume  = {540},
  pages   = {A70},
  year    = {2012}
}

@article{MisnerSharp1964,
  author  = {Misner, C. W. and Sharp, D. H.},
  title   = {Relativistic Equations for Adiabatic, Spherically Symmetric Gravitational Collapse},
  journal = {Phys. Rev.},
  volume  = {136},
  pages   = {B571--B576},
  year    = {1964},
  doi     = {10.1103/PhysRev.136.B571}
}

@article{Cooney2010,
  author  = {Cooney, A. and DeDeo, S. and Psaltis, D.},
  title   = {Neutron stars in $f(R)$ gravity with perturbative constraints},
  journal = {Phys. Rev. D},
  volume  = {82},
  pages   = {064033},
  year    = {2010}
}

@article{deSalas2019,
  author  = {de Salas, P. F. and Malhan, K. and Freese, K. and Hattori, K. and Valluri, M.},
  title   = {On the estimation of the local dark matter density using the rotation curve of the Milky Way},
  journal = {JCAP},
  volume  = {2019},
  pages   = {037},
  year    = {2019}
}

@book{Fliessbach2012,
  author    = {Fliessbach, T.},
  title     = {Allgemeine Relativit\"atstheorie},
  publisher = {Springer},
  edition   = {6},
  year      = {2012}
}

@inproceedings{Bardeen1968,
  author    = {Bardeen, J. M.},
  title     = {Non-singular general-relativistic gravitational collapse},
  booktitle = {Proceedings of the International Conference GR5},
  address   = {Tbilisi, USSR},
  year      = {1968}
}

@article{Hayward2006,
  author  = {Hayward, S. A.},
  title   = {Formation and Evaporation of Nonsingular Black Holes},
  journal = {Phys. Rev. Lett.},
  volume  = {96},
  pages   = {031103},
  year    = {2006}
}

@article{SotiriouFaraoni2010,
  author  = {Sotiriou, T. P. and Faraoni, V.},
  title   = {$f(R)$ theories of gravity},
  journal = {Rev. Mod. Phys.},
  volume  = {82},
  pages   = {451--497},
  year    = {2010}
}

@article{BertoneHooper2018,
  author  = {Bertone, G. and Hooper, D.},
  title   = {History of dark matter},
  journal = {Rev. Mod. Phys.},
  volume  = {90},
  pages   = {045002},
  year    = {2018}
}

@article{GRAVITY2019,
  author  = {{GRAVITY Collaboration} and Abuter, R. and others},
  title   = {Detection of the Schwarzschild precession in the orbit of the star S2 near the Galactic centre massive black hole},
  journal = {Astron. Astrophys.},
  volume  = {636},
  pages   = {L5},
  year    = {2020}
}

@article{Nicolini2009review,
  author  = {Nicolini, P.},
  title   = {Noncommutative black holes, the final appeal to quantum gravity: a review},
  journal = {Int. J. Mod. Phys. A},
  volume  = {24},
  pages   = {1229--1308},
  year    = {2009}
}

@article{Boshkayev2019,
title={A model for a dark matter core at the Galactic Centre},
  author={Boshkayev, Kuantay and Malafarina, Daniele},
  journal={Monthly Notices of the Royal Astronomical Society},
  volume={484},
  number={3},
  pages={3325--3333},
  year={2019},
  publisher={Oxford University Press}
}

@article{Zhao1996,
  author  = {Zhao, H.},
  title   = {Analytical models for galactic nuclei},
  journal = {Mon. Not. R. Astron. Soc.},
  volume  = {278},
  pages   = {488--496},
  year    = {1996}
}

@article{Sharif2016,
  title={{Radiating cylindrical gravitational collapse with structure scalars in $f (R)$ gravity}},
  author={Sharif, M and Yousaf, Z},
  journal={Astrophysics and Space Science},
  volume={357},
  number={1},
  pages={49},
  year={2015},
  publisher={Springer}
}

@article{Astashenok2017,
  author  = {Astashenok, A. V. and Odintsov, S. D. and de la Cruz-Dombriz, A.},
  title   = {The realistic models of relativistic stars in $f(R)=R+\alpha R^2$ gravity},
  journal = {Class. Quantum Gravity},
  volume  = {34},
  pages   = {205008},
  year    = {2017}
}

\end{document}